\documentclass[groupedaddress, showpacs, superscriptaddress]{revtex4-2}
\usepackage{amssymb,amsmath,amsthm,color,graphicx,times,graphicx}
\usepackage{hyperref}
\usepackage{subfigure}
\usepackage{times}
\usepackage{bbold}
\usepackage{color}
\usepackage{array}
\usepackage[french]{babel}
\usepackage{array,multirow,makecell}
\usepackage[table]{xcolor}

\usepackage{orcidlink}
\theoremstyle{plain}

\theoremstyle{definition}

\usepackage{hyperref}
\hypersetup{
	colorlinks=true,
	linkcolor=blue,
	filecolor=blue,      
	urlcolor=blue,
	citecolor=violet,
	pdfauthor={},
	pdftitle={},
	pdfcreator={Abdallah Slaoui},
}
\begin{document}
\title{Improving the probabilistic quantum teleportation efficiency of arbitrary superposed coherent state using multipartite even and odd $j$-spin coherent states as resource}
\author{Meryem El Kirdi}\affiliation{LPHE-Modeling and Simulation, Faculty of Sciences, Mohammed V University in Rabat, Rabat, Morocco.}
\author{Abdallah Slaoui \orcidlink{0000-0002-5284-3240}}\email{Corresponding author: abdallah.slaoui@um5s.net.ma}\affiliation{LPHE-Modeling and Simulation, Faculty of Sciences, Mohammed V University in Rabat, Rabat, Morocco.}\affiliation{Centre of Physics and Mathematics, CPM, Faculty of Sciences, Mohammed V University in Rabat, Rabat, Morocco.}
\author{Hanane El Hadfi}\affiliation{LPHE-Modeling and Simulation, Faculty of Sciences, Mohammed V University in Rabat, Rabat, Morocco.}
\author{Mohammed Daoud}\affiliation{Department of Physics, Faculty of Sciences, University Ibn Tofail, Kenitra, Morocco.}

\begin{abstract}
Quantum teleportation is one of the most important techniques for quantum information secure transmission. Using preshared entanglement, quantum teleportation is designed as a basic key in many quantum information tasks and features prominently in quantum technologies, especially in quantum communication. In this work, we provide a new probabilistic teleportation scheme for arbitrary superposed coherent states by employing the multipartite even and odd $j$-spin coherent states as the entangled resource connecting Alice (sender) and Bob (receiver). Here, Alice possesses both even and odd spin coherent states and makes repeated GHZ states measurements (GHZSMs) on the pair of spins, consisting of ($1$) the unknown spin state and ($2$) one of the two coherent spin states, taken alternately, until reaching a quantum teleportation with maximal average fidelity. We provide the relationship between the entanglement amount of the shared state, quantified by the concurrence, with the teleportation fidelity and the success probability of the teleported target state up to the $n^{\rm th}$ repeated attempt. In this scheme, we show that the perfect quantum teleportation can be done even with a non-maximally entangled state. Furthermore, this repeated GHZSMs attempt process significantly increases both the average fidelity of the teleported state and the probability of a successful run of the probabilistic protocol. Also on our results, we show that the $j$-spin number, the target state parameter and the overlap between coherent states provide important additional control parameters that can be adjusted to maximize the teleportation efficiency.

\vspace{0.25cm}
\end{abstract}
\date{\today}

\maketitle

\section{Introduction}	
Quantum teleportation \cite{Bennett1993} is a physical process via which it is possible to transfer the state of a quantum system from one location to another without knowing the state. In fact, it was one of the most profound discovery of quantum information theory. To accomplish this process, one can use quantum entanglement as shared resources \cite{Einstein1935,Grondalski2002,Li2010}. In this respect, a sender Alice desires to send an unknown state of one qubit to a receiver called Bob across space without having to physically send it. Ideally, Alice and Bob need to share a maximally entangled two qubits state as  resource (quantum channel) and a classical two bits classical channel. Then, Alice performs a Bell state measurement (all the orthogonal Bell states are maximally entangled too) on the combined tripartite state and communicates through the classical channel the result to Bob. Bob, after receiving the result from Alice, performs the suitable unitary transformation to reconstruct the desired state. In this original scheme, the pure maximally entangled state acts as an ideal noise-free quantum channel. Here it is easy to prove that teleportation can succeed faithfully via the maximally entangled state. In other words, the receiver can recover the sent pure state with the probability equal to $1$. This results are extended to $N$-dimensional quantum states \cite{Braunstein2000,Hillery1999}. This protocol is a typical example of an entanglement-assisted process that was first presented by Bennett and coworkers in 1993 \cite{Bennett1993}. Shortly after, this teleportation protocol was experimentally verified in 1997 \cite{Furusawa1998,Boschi1998,Bouwmeester1997}. Since then, various types of quantum teleportation protocols have been widely studied both theoretically \cite{Deng20065,Banaszek2000,Gottesman1999,Yu2014} and experimentally \cite{Kim2001,Takeda2013,Thapliyal2015}.\par

Quantum teleportation protocols enable an unknown quantum state of an object to be transferred from one location to another further away without physically transferring the object itself. Transfer over large distances is necessary for protocols such as quantum networks and quantum computing. Practically, the distances for using optical fibers and free-space channels are restricted to about $100km$. Recently, Ren et al.\cite{Ren2017} reported the quantum teleportation of an independent single-photon qubit from a ground-based observatory to a satellite in low Earth orbit, via an uplink channel, verifying a new record up to $1400km$. Moreover, Wang et al \cite{Wang2015} proved multi-degree-of-freedom quantum teleportation of a single photon encoded in both degrees of freedom as a hyperentangled quantum channel and developed a method to project and discriminate hyper-entangled Bell states. Later, an experimental quantum teleportation of the multi-level state of a single photon in a three-dimensional six-photon system is shown in \cite{Hu2020}. Bell states are of paramount importance in quantum communication and are widely applied in quantum teleportation protocols. Existing Bell state analysis protocols typically focus on encoding the Bell state directly into the physical qubit. Zhou et al \cite{Zhou2015} proposed a comprehensive analysis of the Bell logic state using the helper single atoms in a low-quality cavity and discussed its application in quantum teleportation. Sheng and his colleagues \cite{Sheng2015} described an alternative approach to perform the almost complete logic analysis of the Bell state for the polarized concatenated GHZ state with two logic qubits. The analysis of hyperentangled Bell states is also discussed in \cite{Sheng2010}, where a scheme to fully distinguish 16 hyperentangled Bell states is proposed.\par

Subsequently, several theorists and experiments have proposed, discussed and reported this type of quantum communication. Saliman and his co-workers \cite{Salimian2022} proposed an efficient scheme to teleport an entangled state of two superconducting qubits from Alice's lab to Bob's lab. This kind of two-level system has recently attracted particular attention due to the possibility of tuning the coupling strength between them. They first generated the GHZ state as the chosen quantum channel, and then, by performing the appropriate gates and measurements in each lab, showed that the proposed protocol can be successfully executed with maximum values of fidelity and success probability. Sisodia et al.\cite{Sisodia2022}, provided an optimized controlled quantum teleportation protocol for a multi-qubit quantum state using only one GHZ state with 100$\%$ success probability. Furthermore, Jahanbakhch and his colleagues \cite{Jahanbakhsh2023} presented a teleportation scheme from an unknown atomic state of a qubit (interacting with the quantized field in a cavity) to a second qubit (existing in another distant cavity field) beyond the rotating wave approximation and without the Bell state measurement method. Sisodia at al.\cite{Sisodia2022} investigated an improved scheme for bidirectional quantum teleportation of two-two and two-three qubit quantum states using optimized quantum resource and less consumption of classical resources, where they found improved intrinsic efficiency and discussed the security of the protocol. Moreover, the exact dynamics of entanglement of two two-level atoms in a dissipative cavity and the entanglement protection approach in interacting two-level systems were investigated in \cite{Nourmandipour2015}. Zhou et al.\cite{Zhou2019} proposed bidirectional controlled quantum teleportation of a three-qubit state using GHZ entangled states, where Alice transmits an unknown three-qubit entangled state to Bob, and Bob transmits an unknown three-qubit entangled state to Alice via the control of the supervisor Charlie.\par

In realistic conditions, the impact of the environment exists and the spyder-environment coupling destroys the entanglement existing in the used maximally entangled state \cite{Van2001,Kirdi2023}. The condition of a maximally entangled quantum channel linking Alice and Bob is really hard to achieve or to maintain in its practical implementations due to the noise that reduces entanglement. Therefore, achieving the perfect quantum teleportation will be impossible. The standard figure of merit for quantum teleportation is given by the fidelity deviation \cite{Prakash2013}, which has been the subject of special attention recently. This important quantity gives the idea of fluctuations associated with teleportation fidelity \cite{Badziag2000}. Hence one should consider situations with non maximally entangled state (NME). With a non-maximally entangled quantum channel, the fidelity of teleportation is always less than $1$, so there are two available possibilities. First is to accept imperfect quantum teleportation with a maximal fidelity less than $1$. The second possibility is to get perfect quantum teleportation with probabilities success which is called Probabilistic Quantum Teleportation (PQT) in the literature \cite{Prakash2021}. Indeed, the idea of probabilistic quantum teleportation was first proposed by Agrawal and Pati in \cite{Agrawal2002}, viewed as a generalized quantum teleportation with a non maximally entangled resource and they have shown that one can teleport an arbitrary state with unit fidelity but less than a unit probability. Furthermore, with the choice of probabilistic quantum teleportation, one has reliable quantum teleportation in two cases among four possible results with success probability goes from zero for an unentangled resource state to one-half for the maximally entangled state. The other one-half will come when the shared resource and joint measurement are maximally entangled states.\par

Practically, the number of measurement repetitions depends on the parameter of the shared state, and it ranges from $1$ for a maximally entangled state to infinity for an unentangled state. It should be noted that, in the probabilistic quantum teleportation, both Alice and Bob have to know the shared state. Otherwise, Bob cannot know what is the basis used by Alice to perform its task after receiving the classical communication. Furthermore, quantum teleportation via a partially entangled  shared state was considered in \cite{Li2000}, since a mixed state can be purified to a maximally entangled Bell state \cite{Linden1998,Werner1989,Bennett1996}. So, a mixed quantum channel could never give a reliable teleportation. On the other hand, if one selects an exact teleportation even with some probability, then one should use pure entangled pairs. Further, Fortes and Rigolin \cite{Fortes2016} showed a way to increase the fidelity of teleportation in the presence of noise, without decreasing the probability of success of a protocol execution, working only with probabilistic protocols. They showed that in some cases we cannot obtain a reliable quantum teleportation without using probabilistic protocols. The same team studied the probabilistic quantum teleportation when the entanglement needed to execute the protocol is given by the thermal entanglement \cite{Fortes2017}.\par

In this paper, we shall consider quantum teleportation of spin coherent states. Especially, we shall exploit the factorization property of $su(2)$ coherent states introduced by Peremolov \cite{Perelomov1972,Perelomov1986} which describe many quantum systems with powerful applications. We focus on the probabilistic quantum teleportation using multipartite even and odd $su(2)$ coherent states as a quantum channel. In fact, we shall employ the property according to which even and odd spin coherent states can contain two, three or more spin subsystems \cite{DaoudW2013,Ali2000}. The idea is to teleport the superposed $j$-spin coherent states. We propose here a scheme involving repeated GHZ states measurements to improve both the average fidelity and the success probability of PQT \cite{Zukowski1998,Greenberger1990}. Here, Alice keeps up with the two subsystems on its possession and performs repeated GHZ states measurements on the pair of systems, consisting of $(1)$ the subsystem encoding the information she needs to send and $(2)$ one of both spin of the entangled resource until reaching quantum teleportation with unit maximal fidelity. If perfect quantum teleportation is not attained, i.e., fidelity less than $1$, Alice repeats GHZ states measurement with the entangled subsystem not used in the previous GHZ states measurement replacing the one used. If perfect quantum teleportation is attained, Alice sends the subsystem not used in the last GHZ states measurement and the result of last GHZ states measurements to Bob, who then applies the suitable unitary transformations on his subsystem to get the exact information.\par 

The remaining of the paper is organized as follows. In Section \ref{SC2}, we present a brief review of the multipartite even and odd spin coherent state that we use in the quantum teleportation protocol and discuss their entanglement degree using the concurrence concept in terms of the overlap parameter $p$ and the spin number $j$. In Section \ref{sec3}, we give a detailed description of our probabilistic quantum teleportation scheme, in which perfect transmission of the superposed coherent state is achieved with high efficiency. The primary attempt of the quantum teleportation of spin-$j$ coherent state using GHZ states measurements is reported in Section \ref{Sec4}. Then, we present successively the first and second repeated attempt of GHZ states measurements on the failure cases and we calculate the average fidelity and the success probability of the both attempts. This is done by analyzing their variations in terms of the different parameters characterizing the used teleported quantum state. Finally, we summarize our work in the last section.

\section{Entangled spin coherent states and qubit mapping}\label{SC2}	
\subsection{Even and odd multipartite $j$-spin coherent states}
Broadly, coherent states (quasi-classical states associated to the Heisenberg–Weyl group) are regarded as the closest quantum counterpart to the classical radiation field states, as they are related to the classical wave characteristics of light \cite{Monroe1996,Zhang1990}. Indeed, these states became very important in quantum optics thanks to Glauber \cite{Glauber1963} who proved that they are the eigenstates of the annihilation operator of the harmonic oscillator and minimizing the Heisenberg uncertainty relation \cite{Klauder1985}. Another concept widely used and applied in various quantum information processing and transmission tasks is the notion of spin coherent states introduced by Perelomov \cite{Perelomov1972} (or $su(2)$ coherent states), considered as the quantum states closest to the Glauber's coherent states. These states, also known as atomic coherent states, are non-orthogonal and their qubit states correspond to the spin-$1/2$ representations of $su(2)$. Here, we shall strictly focus on the basic definition of odd and even $su(2)$ coherent state. In particular, by using the property according to which a spin-$j$ coherent state $\left|j, \eta\right\rangle$ can be factorized as a tensor product of two $su(2)$ coherent states $\left|j, \eta\right\rangle\equiv\left|j_{1}, \eta\right\rangle\otimes\left|j_{2},\eta\right\rangle$ with $j=j_{1} + j_{2}$, it is possible to construct a picture where even and odd spin coherent states might be viewed as superpositions of two or more spin coherent systems.\par 
Firstly, the $su(2)$ generators $J_{\pm}$ and $J_{Z}$ satisfy the following structure relations
\begin{equation}
	[J_{Z} , J_{\pm}] = \pm J_{\pm}, \hspace{2cm} [J_{-} ,
	J_{+}] = -2 J_{Z}, 
\end{equation}
where $J_{+}$ and $J_{-}$ represent the raising and lowering operators of $su(2)$ Lie algebra, respectively. The $(2j+1)$-dimensional Hilbert space $\mathcal{H}_{j} $ is spanned by the irreducible tonsorial set $\left\lbrace\vert j,m\rangle, m=-j,-j+1, \cdots, j-1, j\right\rbrace$ characterizing the spin-$j$ representations of the group $su(2)$. On this basis $\left\lbrace\vert j,m\rangle\right\rbrace$, the $su(2)$ generators acting on this irreducible unitary representation as follows
\begin{equation}
J_{\pm}\left|j , m\right\rangle=\sqrt{\left(j\mp m \right)\left(j\pm m+1 \right)}\left|j , m\pm1\right\rangle, \hspace{1cm}J_{Z}\left|j , m\right\rangle=m\left|j , m\right\rangle.
\end{equation}
The coherent state can be obtained by the action of the displacement operator $D(\xi) = \exp (\xi J_{+} - \xi^{\ast} J_{-})$ on the extremal state $|j,-j\rangle$ as
\begin{equation}
	|j,\eta\rangle = D(\xi) |j,-j\rangle = \exp(\xi J_{+} - \xi^{\ast}
	J_{-}) |j,-j\rangle = (1+|\eta|^{2})^{-j} \exp(\eta J_{+})
	|j,-j\rangle \ ,   \label{4.7}
\end{equation}
with $\eta= (\xi/|\xi|)\tan |\xi|$. The even and odd coherent states associated with the $j$-spin are defined by
\begin{equation}
	\left|j,\eta,m\right> = N_m \left(\left|j,\eta\right> + e^{i m \pi}\left|j,-\eta\right>\right),\label{jspincoherentstate}
\end{equation}
where the integer $m$ takes the values $m=0$ (mod $2$) and $m=1$ (mod $1$), $j$ is the quantum angular momentum which takes integer or half integer values and the spin coherent state $\left|j,\eta\right>$ is reduced to
\begin{equation}
	\left|j,\eta\right> = (1 + |\eta|^2 )^{-j} \sum^{j}_{m=-j} \left[ \dfrac{(2j)!}{(j+m)!(j-m)!}\right] ^2\eta ^{j + m} \left|j,m\right>. \label{spincoherentstate}
\end{equation}
For a particle with spin-$\frac{1}{2}$, the above equation (\ref{spincoherentstate}) becomes
\begin{equation}
	|  \eta \rangle = \frac{1}{\sqrt{1 +  |\eta|^{2}}}
	|\downarrow\rangle
	+ \frac{\eta}{\sqrt{1 + |\eta|^{2}}} |\uparrow\rangle . \label{coh}
\end{equation}
We set $\eta $ here, the short for spin-$\frac{1}{2}$ coherent state $|{\frac{1}{2}}, \eta\rangle$ with the basis states
$|\uparrow\rangle \equiv |\frac{1}{2},\frac{1}{2}\rangle$ and
$|\downarrow \rangle \equiv |\frac{1}{2} ,-\frac{1}{2}\rangle$. In this context, any spin-$j$ coherent states can be rewriting as a $2j$ tensorial product of spin-$\frac{1}{2}$ coherent states \cite{Perelomov1986,DaoudW2013}
\begin{align}
	\left|j,\eta\right> = \left(\right|1/2,\eta\left>\right)^{\otimes 2j}\equiv\left(\right|\eta\left> \right)^{\otimes 2j}.\label{3}
\end{align}
The overlap between the coherent states $|j,\eta>$	and $|j,- \eta>$ writes as
\begin{equation}
	p^{2j} =\left\langle j,\eta|j,-\eta\right\rangle= (<\eta|-\eta>)^{2j} = \left( \dfrac{1-|\eta|^2 }{1 + |\eta|^2}\right)  ^{2j},
\end{equation}
and the normalization $N_m$ in the equation (\ref{jspincoherentstate}) is given by	
\begin{equation}
	N_{m} = \left(2 + 2 p^{2j} \cos m \pi\right)^{-\frac{1}{2}}.
\end{equation}
Two asymptotic limits of spin coherent states arise when $p \rightarrow 0 $ and $p \rightarrow 1$. In the first limiting case where $p \rightarrow 0$, the logical qubits $ \vert j, \eta , 0 \rangle$ (even)
and $ \vert j, \eta , 1 \rangle$ (odd) spin coherent states behave
like a multipartite state of Greenberger-Horne-Zeilinger (${\rm GHZ}$) type \cite{Greenberger1990}. In this case, the states $|\eta \rangle $ and $|-\eta \rangle $ are orthogonal and an orthogonal basis can
be defined such that $\vert {\bf 0}\rangle\equiv \vert \eta \rangle$
and $\vert{\bf 1}\rangle \equiv \vert  -\eta \rangle$. Then, the state $\vert j , \eta, m \rangle$ behave like a state of ${\rm GHZ}$-type
\begin{equation}
	\vert j , \eta, m \rangle \sim \vert {\rm GHZ}\rangle_{2j} = \frac
	1{\sqrt{2}}\left(\left| {\bf 0}\right\rangle \otimes \left|{\bf 0}\right\rangle \otimes
	\cdots \otimes\left|  {\bf 0}\right\rangle
	+e^{i m \pi}\left|  {\bf 1}\right\rangle \otimes
	\left| {\bf 1}\right\rangle \otimes \cdots \otimes
	\left| {\bf 1}\right\rangle \right).\label{GHZ}
\end{equation}
In the second limiting case where $p\rightarrow 1$ (or $\eta \rightarrow 0$), it is important to distinguish separately the situations when $m = 0$ (mod $2$) and $m = 1$ (mod $1$). Accordingly, the state $\vert j , \eta, m = 0 ~({\rm mod}~
2)\rangle$ reduces to ground state of the form
\begin{equation}
	\vert j,  0 , 0 ~({\rm mod}~ 2) \rangle \sim  \vert \downarrow
	\rangle \otimes\vert \downarrow \rangle \otimes \cdots \otimes \vert
	\downarrow \rangle,\label{groundstate}
\end{equation}
and for the second situation with odd $m$, the state $\vert j , \eta ,1 ~({\rm mod}~ 1) \rangle$ becomes a multipartite state of $W$-type \cite{Dur2000}
\begin{equation}
	\vert j,  0 ,  1 ~({\rm mod}~ 2) \rangle \sim \vert\text{\rm
		W}\rangle_{2j}
	= \frac{1}{\sqrt{2j}}(\vert \uparrow \rangle \otimes\vert \downarrow \rangle \otimes \cdots\otimes
	\vert \downarrow \rangle  +\vert \downarrow \rangle \otimes\vert \uparrow\rangle \otimes\ldots\otimes \vert \downarrow\rangle
	+\cdots
	+ \vert  \downarrow \rangle \otimes\vert \downarrow \rangle  \otimes \cdots\otimes \vert \uparrow\rangle)~.
	\label{Wstate}
\end{equation}
Also, it is interesting to note that even spin coherent states  $\left| j, \eta , m = 0 ~({\rm mod}~2)\right\rangle $ interpolate between the states of ${\rm GHZ}_{2j}$
$(p \rightarrow 0)$ and the separable state $\vert
\downarrow \rangle \otimes\vert \downarrow \rangle \otimes \cdots
\otimes \vert \downarrow \rangle$ $(p \rightarrow 1)$. Alternatively, the odd case corresponding to $\vert j, \eta , m = 0 ~({\rm mod}~2)
\rangle$ can be  observed as interpolating between ${\rm GHZ}$ ($p \rightarrow 0$) and ${\rm W}$ ($p\rightarrow 1$) type states. The physical advantage of these states is that are the robust states against the decoherence noises \cite{Zurek2003,Slaoui2019,Shaukat2020}. They are widely used and applied to implement and to achieve various quantum information processing tasks. Henceforth, due to the factorization property of $su(2)$ coherent states (\ref{3}), the state (\ref{spincoherentstate}) can also be expressed as 
\begin{align}
	\left|j,\eta,m\right>_{AB} &= N_{m}\left(\right|l,\eta\left>_{A}\otimes\right|j-l,\eta\left>_{B} + e^{im\pi}\right|l,-\eta\left>_{A}\otimes\right|j-l,-\eta\left>_{B}\right)\notag\\&=N_{m}\left(\right|\eta\left>_{A}^{\otimes2l}\otimes\right|\eta\left>_{B}^{\otimes2(j-l)} + e^{im\pi}\right|-\eta\left>_{A}^{\otimes2l}\otimes\right|-\eta\left>_{B}^{\otimes2(j-l)}\right),\label{6}
\end{align}
where $l=0,1/2,1,...,j-1/2,j$. In the following, we shall exploit this factorization property in a teleportation protocol involving multipartite spin coherent states. For this we will consider the expression (\ref{6}), where $A$ denotes the subsystem belonging to Alice and $B$ denotes one of Bob.

\subsection{Target state: Input state (the message)}
By employing a state of the form (\ref{6}) as the quantum channel for quantum communication between Alice and Bob, the input qubit that is teleported from Alice to Bob is assumed to be an arbitrary superposition coherent state. It is given by
\begin{equation}
	\left|I\right>_{C} = a \left|j-l, \eta\right>_{C} + b \left|j-l, - \eta\right>_{C}\equiv a \left|\eta\right>_{C}^{\otimes2(j-l)} + b \left|- \eta\right>_{C}^{\otimes2(j-l)},\label{InputState}
\end{equation}
where $|a|^{2}+|b|^{2}=1$ and $C$ denotes the spin coherent state at hand of a third party called Charlie. Using the notation adopted in reference \cite{DaoudW2013}, this subsystem can be rewritten in the orthogonal basis $\{ \left|0\right>_{(j-l)}, \left|1\right>_{(j-l)} \}$  defined as		
\begin{align}
	\left|0\right>_{(j-l)} = \dfrac{\left|\eta\right>^{\otimes2(j-l)} + \left| -\eta\right>^{\otimes2(j-l)}}{\sqrt{2(1+p^{2(j-l)})}},\hspace{2cm}\left|1\right>_{(j-l)} = \dfrac{\left|\eta\right>^{\otimes2(j-l)} - \left| -\eta\right>^{\otimes2(j-l)}}{\sqrt{2(1-p^{2(j-l)})}},
\end{align}
and the state (\ref{InputState}) becomes
\begin{equation}
	\left|I\right>_{C} =\textit{N}\left( A_{j-l}(a+b) \left|0\right>_{(j-l)} + B_{j-l}(a-b) \left|1\right>_{(j-l)}\right),\label{9}
\end{equation}				
where 	\begin{equation}
	A_{j-l} = \sqrt{\dfrac{1 + p^{2(j-l)}}{2}}, \hspace*{1cm}  B_{j-l} = \sqrt{\dfrac{1 - p^{2(j-l)}}{2}},\hspace*{1cm} N = (1+2abp^{2(j-l)})^{-1/2},
\end{equation}
involving the overlap $p$ which is related to the non-orthogonality of two spin coherent states of equal amplitude and opposite phase. In this scheme, the state $\left|j,\eta,m\right>$ can be expressed as a state of two logical qubits
\begin{align}	
\left|j,\eta,m\right> = &N_{m}(A_{l}A_{j-l} \beta_{+}\left|0\right>_A^{\otimes2l}\left|0\right>_{B}^{\otimes2(j-l)}+A_lB_{{j-l}}\beta_{-}\left|0\right>_A^{\otimes2(j-l)}\left|1\right>_{B}^{\otimes2(j-l)}\notag\\&
+ A_{j-l}B_{l} \beta_{-}\left|0\right>_{A}^{\otimes2(j-l)}\left|1\right>_B^{\otimes2(j-l)}+B_{l}B_{j-l} \beta_{+}\left|1\right>_A^{\otimes2l}\left|1\right>_{B}^{\otimes2(j-l)}). 
\end{align}
with $\beta_{\pm} = 1 \pm e^{im \pi}$. In probabilistic quantum teleportation, which is based on the repeating quantum measurements over the parts that constitute the quantum channel, it is necessary that both Alice and Bob have the same $j$-spin number. For this purpose, we consider the splitting $l=j-l=\frac{j}{2}$ which arises from the decomposition of even and odd coherent states associated with the spin $j_{1}+j_{2}=j$. Then, the state $\left|j,\eta,m\right>$ writes	as
\begin{align}	
\left|j,\eta,m\right> = &N_{m}(A_{j/2}A_{j/2} \beta_{+}\left|0\right>^{\otimes j}_A\left|0\right>^{\otimes j}_{B} + A_{j/2}B_{{j/2}} \beta_{-}\left|0\right>^{\otimes j}_A\left|1\right>^{\otimes j}_{B}\nonumber \\
	& + A_{j/2}B_{j/2} \beta_{-}\left|0\right>^{\otimes j}_{A}\left|1\right>^{\otimes j}_B + B_{j/2}B_{j/2} \beta_{+}\left|1\right>^{\otimes j}_A\left|1\right>^{\otimes j}_{B}),\label{13}
\end{align}
In fact, an odd or even spin-$j$ coherent state holds an amount of intrinsic entanglement between its parts due to the division of the $j$-spin into two or more sub-parts. This can be understood as the existence of a single-particle entanglement that is emphasized by the authors of several works \cite{Binicioglu2007,Can2005,Terra2007}. To examine the influence of this intrinsic entanglement on our probabilistic quantum teleportation protocol, we can use Wootters concurrence which is a most widely accepted entanglement measure for a two-party system \cite{Wootters2001}. Sure enough, it can be extended and adapted to multipartite coherent states. For the state under consideration $	\left|j,\eta,m\right>$ (\ref{13}), the amount of entanglement is given by the following expression
\begin{align}
	{\cal C}\left(\left|j,\eta,m \right>\right) & = \dfrac{1 - p^{2j}}{1 + p^{2j}cos m\pi}.\label{14}
\end{align}
\begin{figure}[h!]	\centering
	\includegraphics[scale=0.7]{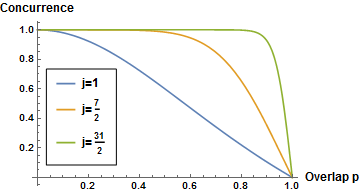}
	\caption{Variation of the concurrence versus the overlap $p$ for $m=0$.}\label{Fig1}
\end{figure}
The behavior of the concurrence ${\cal C}\left( \left|j,\eta,m \right>\right) $ for even spin coherent states versus the overlap $p$ is plotted in the Fig.(\ref{Fig1}). It is shown that the degree of quantum correlations necessarily depends on the $j$-spin number. We can see that the concurrence increases with the increase of $j$-spin number, e.g., the entanglement for $j=31/2$ is greater than other values of spin $j$ and for any value of the overlap $p$. Also, for a given spin $j$, the maximal value of ${\cal C}$ is reached in the bipartition where $j_{1}=j_{2}=j/2$. Furthermore, for $p\rightarrow 0$, the state (\ref{13}) reduces to a bipartite state of GHZ-type which is maximally entangled (${\cal C}=1$). Conversely, the even spin coherent state is separable when $p\rightarrow1$. In fact, we have ${\cal C}=0$ for $m$ even (i.e., symmetric pure states). Contrary to the later case, for odd spin coherent states with $m=1$ (mod $1$), we have ${\cal C}\left(\left|j,\eta,m \right>\right)=1$ as it can be verified from the expression (\ref{14}). In this case, the states (\ref{13}) are maximally entangled states and include GHZ (\ref{GHZ}) and W (\ref{Wstate}) type states.

\section{Spin Coherent States in Probabilistic Teleportation Protocol}\label{sec3}
The main objective of this technique is to teleport a state from $A$ to $B$ by sending two bits through a classical way. In the following, we describe a simple scheme to teleport an unknown spin state of the form Eq.(\ref{InputState}) from Alice to Bob using a spin coherent state. In particular, we consider the GHZ states measurements on a pair of spin to improve success probability of the quantum teleportation. Alice and Bob initially share a state of the form Eq.(\ref{13}) and apply the following steps:
\begin{description}
	\item[($i$)] Alice has initially in its possession both even and odd spin coherent state and she performs the GHZSMs on the pair of particles in her possession $1$ (the unknown state of the particle to be teleported) and $2$ (a part of the shared two spin coherent state taken alternatly). If perfect quantum teleportation is achieved, then she perform to the next steps.
	\item[($ii$)] Alice will send the spin used in the last GHZSMs and the result of her last  measurement to Bob by a classical way.
	\item[($iii$)] Depending on the result of Alice's GHZSMs, Bob performs a unitary transformation allowing him to find the target state (state to be teleported). If perfect protocol is not attained, Alice repeats GHZSMs on a pair of particle states, $1$ (the state to be teleported) and $2$ (one of the two coherent state spin not used in the previous measure).
\end{description} 
Once Bob get the desired state, we evaluate the degree of fidelity of the state by comparing it with the starting state. In probabilistic quantum teleportation, Alice finds a success in only two cases (cases with unit maximum fidelity) and then, the probability of success will be the sum of the two probabilities for the two cases corresponding to a successful attempt. For the two cases where we find failure, Alice can accept half-success probability or, in order to get high success probability, she can continue with the exchange of the spins of the shared resource state, and the repetition of the GHZSM, which is called probabilistic quantum teleportation by repetitions.
\begin{figure}[h!]	\centering
	\includegraphics[scale=0.65]{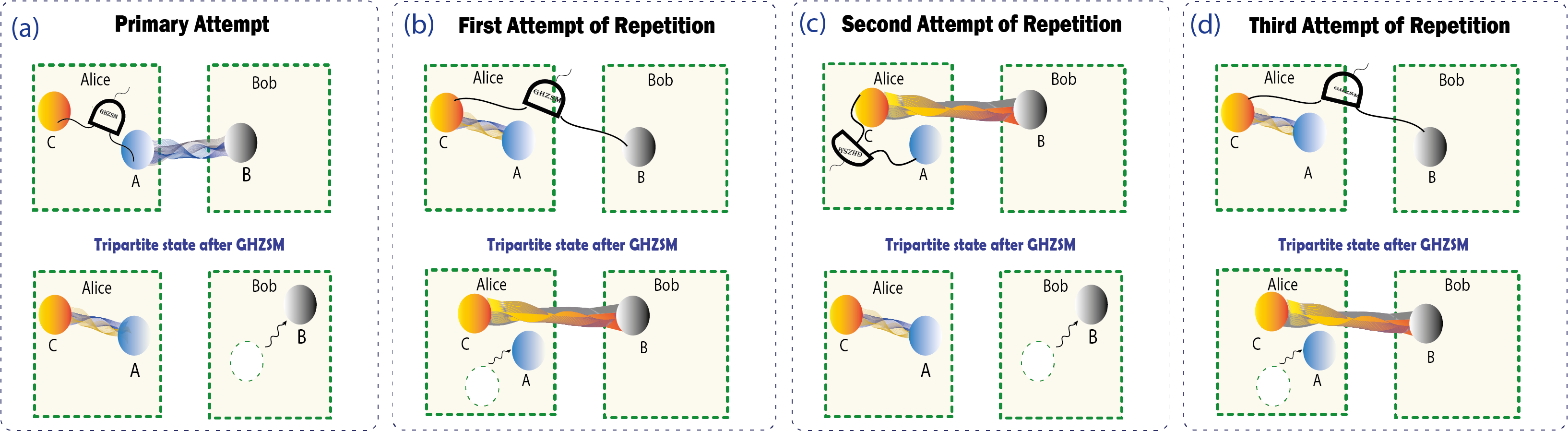}
	\caption{Measurement repetition process by GHZ states. Repetition takes place when failure is indicated.}
\end{figure}

In the measurement process, each GHZ states measurements attempt indicates some results with probabilistic success and others with failure. For cases with an indicated success, Alice sends the particle not used in the GHZ state measurements to Bob, who then applies an appropriate unitary transformation to exactly reproduce the unknown information state. For the cases with failure indicated, another GHZ states measurements can be made but with the particle A and the particle not used during the previous attempt, because repeating GHZ states measurements with the same particles will give the same results, so nothing new. For primary attempt, Alice initially has a pair of particles $C$ and $A$, where $C$ is the particle to be teleported and $A$ is a part of the shared entangled resource state between Alice and Bob, then Alice performs GHZ states measurements on particles in its possession ($C$ and $A$). After GHZSMs, the output state of particle $B$ is found and making particles $C$ and $A$ entangled (Panel $2.(a)$). Practically in this primary attempt, success is indicated in two cases and failure in two cases. For the cases with indicated success, Alice sends Bob the unused particle and completes her task. For the cases with indicated failure, Alice repeat GHZ states measurements.\par

The first repeated attempt of GHZSMs will be done when Alice faced with failure in the primary attempt of GHZSMs, i.e., the maximum fidelity is less than unity. At this step, the particles entangled are $C$ and $A$, so Alice can perform the GHZSMs on particles $C$ and $B$, finding the out state of particle $A$, and making particles $C$ and $B$ entangled (Panel $2.(b)$). Practically in this first attempt of repetition, success is found in four cases and failure in others four cases. The second repeated attempt of GHZSMs can be performed when Alice find failure in the first repeated attempt. At this level, Alice can apply repeated GHZSMs on particles $C$ and $A$, making them again entangled and getting the state of particle $B$ (Panel $2.(c)$). The third repeated attempt of GHZSMs will be applied when Alice finds that the second repeated GHZ measurement attempt fails, in this attempt and after the GHZSMs, Alice will find the state of particle $A$, which means that particles $C$ and $B$ are now entangled (Panel $2.(d)$). In principle, this repeating process can be continued until success is indicated.

\section{Entanglement Swapping and repeated GHZ states measurements}\label{Sec4}	
To teleport a state of the form (\ref{InputState}), we consider the combined tripartite state
\begin{equation}
	\left|\psi\right>_{ABC} = \left|I\right>_{C} \otimes \left|j,\eta,m\right>_{AB},
\end{equation}
where Alice keeps all the spins with her and makes GHZ states measurements on subsystems $A$ and $B$ which respectively contained the two spin coherent states of the entangled resource.
\subsection{Primary attempt of GHZ states measurements}								
Alice chooses the following generalized orthogonal $GHZ$ states \cite{Wu2000} basis spanned by the four vectors	
\begin{align}
	\left|{\cal PA}\right>_{0}^{CA}\equiv\left|GHZ\right>_{0} & = \dfrac{1}{\sqrt{2}}\left(\left|0\right>^{\otimes j}_{C} \otimes
	\left|0\right> ^{\otimes j}_{A}
	+ \left|1\right>^{\otimes j}_{C}  \otimes
	\left| 1\right>^{\otimes j}_{A} \right), \hspace{1cm}
	\left|{\cal PA}\right>_{1}^{CA}\equiv\left|GHZ\right>_{1} = \dfrac{1}{\sqrt{2}}\left(\left|0\right>^{\otimes j}_{C} \otimes
	\left|1\right> ^{\otimes j}_{A}
	+ \left|1\right>^{\otimes j}_{C}  \otimes
	\left| 0\right>^{\otimes j}_{A} \right), \nonumber \\
	\left|{\cal PA}\right>_{2}^{CA}\equiv\left|GHZ\right>_{2} & = \dfrac{1}{\sqrt{2}}\left(\left|0\right>^{\otimes j}_{C} \otimes
	\left|1\right> ^{\otimes j}_{A}
	- \left|1\right>^{\otimes j}_{C}  \otimes
	\left| 0\right>^{\otimes j}_{A}\right), \hspace{1cm}
	\left|{\cal PA}\right>_{3}^{CA}\equiv\left|GHZ\right>_{3} = \dfrac{1}{\sqrt{2}}\left(\left|0\right>^{\otimes j}_{C} \otimes
	\left|0\right> ^{\otimes j}_{A}
	- \left|1\right>^{\otimes j}_{C}  \otimes
	\left| 1\right>^{\otimes j}_{A}\right),
\end{align}	
and performs GHZ-states measurements on the pair of subsystems $AC$. After measurements, the corresponding states of the particle $C$ become $\left|T_{i}\right>$ for GHZSMs result $i$ (with $i=0,1,2,3$). If the quantum teleportation is perfect, Alice sends GHZ measurement result $\left|T_{i}\right>$ to Bob (see Table.(\ref{tab1})). Otherwise, she performs a new GHZ states measurements. In this picture, the probabilities $\mathcal{P}_{i}$ are respectively given by:
\begin{align}
	\mathcal{P}_{0} = \dfrac{\left(NN_{m}\right) ^2}{2}\left[A_{j/2}\left(A_{j/2}^2(a+b)\beta_{+} + B_{j/2}^{2}\left(a-b\right)\beta_{-} \right)^2+ B_{j/2}\left(A_{j/2}^2\left( a+b\right)\beta_{-} +  B_{j/2}^{2}\left(a-b\right)\beta_{+}\right)^2\right]\nonumber,
\end{align}
\begin{align}
	\mathcal{P}_{1} = \dfrac{\left(NN_{m}\right)^2}{2}\left[B_{j/2}A_{j/2}^{2} \left( \left(a+b\right)\beta_{-} +\left(a-b\right)\beta_{+} \right)^{2}
	+A_{j/2}B_{j/2}^2\left(\left(a+b\right)\beta_{+} + \left(a-b\right) \beta_{-}\right)^2\right]\nonumber,
\end{align}
\begin{align}
	\mathcal{P}_{2} = \dfrac{(NN_m)^2}{2}\left[B_{j/2}A_{j/2}^{2} \left(\left(a+b\right) \beta_{-} -\left(a-b\right)\beta_{+} \right)^{2} +A_{j/2}B_{j/2}^{2} \left(\left(a+b\right)\beta_{+} -\left(a-b\right)\beta_{-} \right)^{2}\right]\nonumber,
\end{align}
\begin{align}
	\mathcal{P}_{3} = \dfrac{(NN_m)^2}{2}\left[
	A_{j/2}	\left(A_{j/2}^{2}\left(a+b\right)\beta_{+}- B_{j/2}^{2}\left(a-b\right)\beta_{-} \right)^{2}+
	B_{j/2}	\left(A_{j/2}^{2}(a+b)\beta_{-} - B_{j/2}^{2}\left(a-b\right) \beta_{+}\right)^{2}\right].
\end{align}	
To do better than classical teleportation scheme, we need the shared quantum state to be entangled. Furthermore, if two resource states are used, then it is necessary to find which resource state is the better one to teleport an unknown quantum state. In this context, the quantum teleportation efficiency through noisy quantum channels is quantified by the quantum fidelity. It is a measure of the overlap between a state to be teleported (\ref{InputState}) and a teleported state $\left|T_{i}\right\rangle$ and was defined in \cite{Bennett1993} as follows
\begin{equation}
	\mathcal{F}_i = |\left<T_{i}|I\right>_{C}|^2,
\end{equation}
with $\left|I\right>_{C}$ is the input state and $\left|T_{i}\right>$ is the output state. Therefore, one gets
\begin{align}
	\mathcal{F}_0 & = \dfrac{(N^2N_m)^2}{2p_0}|(A_{j/2}^2(a+b) +B_{j/2}^2(a-b))^2 + (A_{j/2}^2(a+b) -B_{j/2}^2(a-b))^2e^{im\pi}|^2\nonumber,
\end{align}
\begin{align}
	\mathcal{F}_1 & =\dfrac{2(N^2N_m)^2}{p_1} |B_{j/2}A_{j/2}^3(a+b)(a-be^{im\pi})
	+A_{j/2}B_{j/2}^3(a-b)(a+be^{im\pi})|^2\nonumber,
\end{align}
\begin{align}
	\mathcal{F}_2 & =\dfrac{2(N^2N_m)^2}{p_2} |B_{j/2}A_{j/2}^3(a+b)(b-ae^{im\pi})
	+A_{j/2}B_{j/2}^3(a-b)(b+ae^{im\pi})|^2\nonumber,
\end{align}		
\begin{align}
	\mathcal{F}_3 & = \dfrac{(N^2N_m)^2}{2p_3}|(A_{j/2}^4(a+b)^2 -B_{j/2}^4(a-b)^2)(1+e^{im\pi}) |^2,
\end{align}
and the average fidelity writes as
\begin{equation}\label{faveragF}
	\mathcal{F}_{av}  =\sum_i \mathcal{F}_i \mathcal{P}_i.
\end{equation}			
This quantity characterizes the amount of fluctuations associated with teleportation fidelity \cite{Agrawal2002,Oh2002}. We set $NA_{j/2}(a+b)= \cos(\omega /2)$ and $NB_{j/2}(a-b)= \sin(\omega /2)$, where $\omega \in [0, 2 \pi]$ (look at $\cos^{2}(\omega/2)+\sin^{2}(\omega/2)=1$). For $m=0$ (mod $2$), straightforward calculations leads to
\begin{align}
	\mathcal{F}_{av.2} & = \dfrac{1}{4}\left( \dfrac{|p^j + cos \omega|^2+|1+ p^j cos \omega |^2 }{1+p^{2j}} +sin^2 \omega\right).\label{eq27}
\end{align}
To understand the interplay between the quantum teleportation fidelity and the quantum channel entanglement, we express the average fidelity associated with the primary attempt by GHZ states measurements in terms of the concurrence. Indeed, using the equations (\ref{14}) and (\ref{eq27}), one gets
\begin{align}
	F_{av.2} = \dfrac{1}{8}\left( |\sqrt{1-{\cal C}} +\sqrt{1+{\cal C}} \cos \omega|^2+|\sqrt{1+{\cal C}} +\sqrt{1-{\cal C}}\cos\omega|^2 +2\sin^2 \omega\right),\label{28}
\end{align}
Here, one has to treat separately three cases $\omega=0,\pi/2,\pi$; For $\omega=\pi/2$, i.e. the teleported state is the superposed target state, the average fidelity (\ref{28}) reduces to $F_{av.2} =1/2$. When the teleported state is the separable target state (i.e. $\omega=0$ or $\omega=\pi$), the formula (\ref{28}) reads as
\begin{equation}
	F_{av.2}=\left\lbrace
	\begin{aligned}
		\dfrac{1+\sqrt{1-{\cal C}^2}}{2},\hspace{1cm}{\rm if}\hspace{1cm}\omega=0,\\
		\dfrac{1-\sqrt{1-{\cal C}^2}}{2},\hspace{1cm}{\rm if}\hspace{1cm}\omega=\pi
	\end{aligned}
	\right.
\end{equation}
For antisymmetric spin coherent states with $m=1$ (mod $1$), the average fidelity (\ref{faveragF}) becomes 	
\begin{align}
	\mathcal{F}_{av.1} = \dfrac{1}{4}\left( 1+|cos\omega|^2+sin^2 \omega\right).
\end{align}	
Then, for this primary attempt of GHZSMs (see Table.\ref{tab1}), it is clear that for $m=0$, one has success if the GHZSMs results is associated to $\left|GHZ\right>_{0}$ or $\left|GHZ\right>_{3}$, and for $m=1$, the success is indicated with the results associated to $\left|GHZ\right>_{1}$ or $\left|GHZ\right>_{2}$. Otherwise, the result is associated to $\left|GHZ\right>_{1}$ or $\left|GHZ\right>_{2}$ (for $m=0$) and $\left|GHZ\right>_{0}$ or $\left|GHZ\right>_{3}$ (for $m=1$) in the failure cases. In this attempt, the expression of probability success is the sum of  the probabilities to get the states $\left|GHZ\right\rangle_{0}$ and $\left| GHZ\right\rangle_{3}$, i.e., $\textit{P}_{\textit{success.2}}= \mathcal{P}_0 + \mathcal{P}_3 $. Also, it is important to notice that for the states encompassing $m=1$, the probability success takes the form $\textit{P}_{\textit{success.1}}= \mathcal{P}_{1} + \mathcal{P}_{2} =1/2$. Thus, we obtain the total probability of this successful teleportation as
\begin{align}
	{P}_{success.2} =\textit{P}_{\textit{success.1}} + \dfrac{p^j cos\omega}{1 + p^{2j}}.
\end{align}
\begin{figure}[h!]
	{{\begin{minipage}[b]{.5\linewidth}
				\centering
				\includegraphics[scale=0.6]{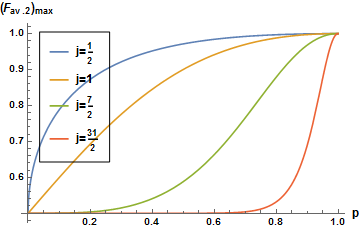}\vfill
				$(a)$
			\end{minipage}\hfill
			\begin{minipage}[b]{.5\linewidth}
				\centering
				\includegraphics[scale=0.6]{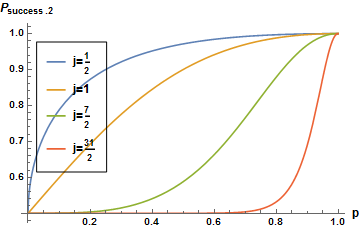}\vfill
				$(b)$
			\end{minipage}}}
	\caption{(a) the maximal average fidelity, (b) the success probability cursus as a function of the overlap parameter $p$ for different values of spin $j$ when $m=0$.}\label{Fig3}
\end{figure}

In Fig.(\ref{Fig3}), we depict the variation of the maximal average fidelity with respect to the overlap $p$ for various values of the $j$-spin. We can clearly see that for $p$ approaching the unity, the maximum average fidelity approaches $1$ too. In order to understand the interpretation of this behavior, we must consider the expression of the degree of entanglement of the initially shared state between Alice and Bob in terms of the overlap parameter $p$. If we go back to equation (\ref{14}) and focus on figure (\ref{Fig1}) which represents the variation of the concurrence versus the overlap $p$ for $m=0$, we can observe that when $p$ approaches $1$, the shared state is non-maximally entangled. This indicates that for $m=0$, the average fidelity reaches its maximal value if the shared state between Alice and Bob is a non-maximally entangled state. On the other hand, we can verify that when the number of spin $j$ increases, the value of the maximum average fidelity decreases until reaching the classical fidelity $1/2$ for a large number of spin $j$.\par
In this way, for $m=0$ and $p=0$, i.e., the shared resource  is a maximally entangled state (GHZ state (\ref{GHZ})), the maximal average fidelity is one-half. For $m=0$ and $0<p<1$, i.e., the shared resource state is a non-maximally entangled state, the maximal average fidelity depends on the number of spin $j$ and it is between $1$ and $1/2$ for an important number of spin $j$. For $m=0$ and $p=1$, i.e., the shared resource is a separable state (ground state (\ref{groundstate})), the maximal average fidelity is $1$ for different value of spin $j$. Now, for $m=1$, the shared resource state is generally a maximally entangled state, for $p=0$ (GHZ state) and for $p=1$ (W state (\ref{Wstate})), about maximal average fidelity is always $1/2$. This proves that non-maximally entangled state can presents more interesting  results in comparison with maximally entangled state.\par

The plot in Fig. (\ref{Fig3}) shows the behavior of the success probability as a function of the overlap $p$ for some specific values of the parameter $\omega$ of the input state ($\omega= 0$ or $\omega=\pi$) for which the average fidelity takes its maximal value. One can observe immediately that, for $m=0$, the success probability depends on the spin number $j$; more than this number increases more than the probability decreases until it will be achieved by its minimal value $1/2$ for a large number of spin $j$.\par

To successfully teleport an arbitrary superposition coherent state of type (\ref{InputState}) by involving the even spin coherent states ($m=0$) as a shared resource, we see clearly that one needs at least $R$ repetitions number to get a quantum teleportation with unit probability, with $R$ is the reciprocal of the success probability given in \cite{Agrawal2002}. In our case we obtain
\begin{align}
	R = \dfrac{4(1 + p^{2j})}{p^{2j}+2 p^j cos\omega+1}.
\end{align}		
For $m=1$, the success probability is independent on the overlap parameter $p$ (since $\textit{P}_{\textit{success.1}}=1/2$) and it takes the value $1/2$ for a maximally entangled state (both for W and GHZ type states). For $m=0$, the success probability goes from zero (when the shared resource state is a ground state ($p=1$) and $\omega=\pi$, then the number of repetitions $R$ tends to infinity) to one-half (when the shared resource state is GHZ state ($p=0$) or when the shared resource state is a GHZ state ($p=1$) and $\omega=\pi/2$, then the minimal number of repetitions is $R=4$ ) to one for a ground shared resource state (when $p=1$ and $\omega =0$ then number of repetitions is at least $R=2$).
	\begin{table}[h!]
		\centering
		\renewcommand{\arraystretch}{1.3}
		\begin{tabular}{|c| c| c| c|c|}
			\hline \rowcolor{lightgray}GHZSM &  State of particle B &  $\mathcal{F}_{max.2}$ &  $\mathcal{F}_{max.1}$ & $\mathcal{P}$\\
			\hline  $\left|GHZ\right>_{0}$ & $\begin{array} {lcl}\left|T_{0}\right>=\dfrac{NN_m}{\sqrt{2p_0}}&(&[A_{j/2}^3(a+b)\beta_{+} + A_{j/2}B_{j/2}^2(a-b)\beta_{-}]	\left| 0\right>^{\otimes j}\\ &+&[A_{j/2}^2B_{j/2}(a+b)\beta_{-}+  B_{j/2}^3(a-b)\beta_{+}]	\left| 1\right>^{\otimes j} )_{B}\end{array}$ &  $1$ &  $\neq 1$&$\mathcal{P}_0$\\
			\hline  $\left|GHZ\right>_{1}$ & $\begin{array} {lcl}\left|T_{1}\right>=\dfrac{NN_m}{\sqrt{2p_1}}&(&B_{j/2}A_{j/2}^2[(a+b)\beta_{-} + (a-b)\beta_{+}]	\left| 0\right>^{\otimes j}\\& +&[B_{j/2}^2A_{j/2}((a+b)\beta_{+} +  (a-b)\beta_{-})]	\left| 1\right>^{\otimes j} )_{B}\end{array}$ & $\neq 1$  &  $1$ & $\mathcal{P}_1$\\
			\hline  $\left|GHZ\right>_{2}$ & $\begin{array} {lcl}\left|T_{2}\right>=\dfrac{NN_m}{\sqrt{2p_2}}&(&B_{j/2}A_{j/2}^2[(a+b)\beta_{-} - (a-b)\beta_{+}]\left| 0\right>^{\otimes j} \\&+& A_{j/2}B_{j/2}^2[(a+b)\beta_{+} -  (a-b)\beta_{-}]\left| 1\right>^{\otimes j})_{B}\end{array}$ &  $\neq 1$ &  $1$ &$\mathcal{P}_2$ \\
			\hline  $\left|GHZ\right>_{3}$ & $\begin{array} {lcl}\left|T_{3}\right>=\dfrac{NN_m}{\sqrt{2p_3}}&(&\left[A_{j/2}^3(a+b)\beta_{+} - A_{j/2}B_{j/2}^2(a-b)\beta_{-}\right]	\left| 0\right>^{\otimes j} \\&+&\left[A_{j/2}^2B_{j/2}(a+b)\beta_{-} -B_{j/2}^3(a-b)\beta_{+}\right]	\left| 1\right>^{\otimes j} )_{B}\end{array}$ &  $1$  &  $\neq 1$&$\mathcal{P}_3$\\
			\hline
		\end{tabular}
		\caption{$GHZ$ states measurements results in primary attempt on particles $A$ and $C$.}
		\label{tab1}
	\end{table}

For indicated success, i.e., the results $\left\lbrace\left|GHZ\right>_{0} {\rm or} \left|GHZ\right>_{3}\right\rbrace$ and  $\left\lbrace \left|GHZ\right>_{1} {\rm or} \left|GHZ\right>_{2}\right\rbrace$ obtained for $m=0$ and $m=1$ respectively, Alice sends to Bob the particle not used in GHZ state measurement (particle $B$) and Bob performs a suitable unitary transformation to get the exact information. If she faced with failure, then she repeats GHZ state measurement but with the systems pair $CB$ not used in the previous GHZ states measurements and not with the pair of $CA$ because a repeated GHZ states measurements with the same pair of particles will not give something new. In the following, Alice will perform new GHZ states measurements on failure cases by using the pair $CB$ of even and odd spin coherent states.
\subsection{First repeated attempt of GHZ states measurements}
To carry out the process achieving the desired success, Alice makes first repeated attempt of GHZ states measurements. If Alice faced with failure in primary attempt of GHZSMs, then she can repeat GHZSMs on pair particles $CB$. For the states with $m=0$, the two cases were GHZSMs results $\left|GHZ\right>_{1}$ and $\left|GHZ\right>_{2}$. We examine here each of these two cases:\par

{\bf Case -1:} In this first case, the input state after primary GHZ states measurements result $\left|GHZ\right>_{1}$ is given by
\begin{align}
	\left|\psi{(1)} \right>_{ABC}^{m=0}=\left|GHZ \right>_{1}\otimes \left|T_{1} \right\rangle = & \dfrac{NN_m}{2\sqrt{p_1}}(\left|0\right>^{\otimes j}_{C}\otimes\left|1\right>^{\otimes j}_{A}+ \left|1\right>^{\otimes j}_{C}\otimes \left|0\right>^{\otimes j}_{A})\otimes(
	B_{j/2}A_{j/2}^2[(a+b)\beta_{-}
	+\nonumber \\
	&(a-b)\beta_{+}]\left|0\right>^{\otimes j}_{B}+
	B_{j/2}^2A_{j/2} [(a+b)\beta_{+} + (a-b)\beta_{-}]\left|1\right>^{\otimes j}_{B}),
\end{align}
Here, Alice performs the $1^{\rm st}$ repeated attempt on the couple of particles ($C$,$B$) and she uses the GHZ orthogonal basis of the subsystems $C$ and $B$ as follows
\begin{align}
	\left|GHZ\right>_{10} =\left|{\cal PA}\right>_{0}^{CB}, \hspace{1cm}
	\left|GHZ\right>_{11}= \left|{\cal PA}\right>_{1}^{CB}, \hspace{1cm}
	\left|GHZ\right>_{12}= \left|{\cal PA}\right>_{2}^{CB},\hspace{1cm}			
	\left|GHZ\right>_{13} = \left|{\cal PA}\right>_{3}^{CB}.
\end{align}
It should be noted that the first digit of states $\left|GHZ\right\rangle_{0,1,...,n^{th}}$ denotes GHZSMs result in primary attempt, second digit denotes $GHZSM$ result in first attempt (see table $1$) and so on $n^{th}$ digit denotes GHZSMs result in $\left(n-1 \right)^{th}$ attempt. The corresponding probabilities are found as
\begin{align}
	\mathcal{P}_{10} = \mathcal{P}_{13}  =  \dfrac{(NN_m)^2}{8p_1}\left[
	\left( A_{j/2} B_{j/2}^2((a+b)\beta_+ +(a-b)\beta_-)\right)^2+ \left( A_{j/2}^2B_{j/2}((a+b)\beta_- +(a-b)\beta_+)\right)^2\right],\nonumber
\end{align}
and 
\begin{align}
	\mathcal{P}_{11} =\mathcal{P}_{12} =\dfrac{(NN_m)^2}{8p_1}\left[
	\left(B_{j/2} A_{j/2}^2((a+b)\beta_- +(a-b)\beta_{+})\right)^2+
	\left(B_{j/2}^2A_{j/2}((a+b)\beta_{+} +(a-b)\beta_{-})\right)^2\right].\nonumber
\end{align}
Then, the corresponding teleportation fidelity are evaluated as
{ {\begin{align}
			\mathcal{F}_{10}= \dfrac{2(N^2N_m)^2}{p_1p_{10}}|A_{j/2}^2B_{j/2}^2(a^2+b^2e^{im\pi}) |^2,\hspace{1cm}\mathcal{F}_{13} =  \dfrac{(N^2N_m)^2}{2p_1p_{13}}|A_{j/2}^2B_{j/2}^2(a^2+b^2)(1+e^{im\pi})|^2.\nonumber
\end{align}}}
{ \textit{	\begin{align}
			\mathcal{F}_{11} =\dfrac{(N^2N_m)^2}{2p_1p_{11}} |B_{j/2}A_{j/2}^3(a+b)(a-be^{im\pi})
			+A_{j/2}B_{j/2}^3(a-b)(a+be^{im\pi})|^2,\nonumber
\end{align}	}}
\begin{align}
	\mathcal{F}_{12}  =  \dfrac{(N^2N_m)^2}{2p_1p_{12}}
	|B_{j/2}A_{j/2}^3(a+b)(a-be^{im\pi})
	-A_{j/2}B_{j/2}^3(a-b)(b+ae^{im\pi})|^2,\nonumber
\end{align}

{\bf Case -2:} The input state after primary $GHZ$ states measurements result $\left|GHZ_{2}\right>$ is:
\begin{align}
	\left|\psi{(2)}\right >_{ABC}^{m=0}=\left|GHZ \right>_{2}\otimes \left|T_{2} \right\rangle	 = \dfrac{NN_m}{2\sqrt{p_2}}(&\left|0\right>^{\otimes j}_{C}\otimes\left|1\right>^{\otimes j}_{A} - \left|1\right>^{\otimes j}_{C}\otimes \left|0\right>^{\otimes j}_{A})\otimes(	B_{j/2}A_{j/2}^2[(a+b)\beta_{-}\nonumber\\
	& -(a-b)\beta_{+}]\left|0\right>^{\otimes j}_{B}+B_{j/2}^2A_{j/2} [(a+b)\beta_+ - (a-b)\beta_-]\left|1\right>^{\otimes j}_{B}),
\end{align}
and Alice may choose orthogonal GHZ basis of subsystems $C$ and $B $ as
\begin{align}
	\left	|GHZ\right> _{20}= \left|{\cal PA}\right>_{0}^{CB}, \hspace{1cm}
	\left|GHZ\right>_{21}  =\left|{\cal PA}\right>_{1}^{CB}, \hspace{1cm}
	\left|GHZ\right>_{22} =\left|{\cal PA}\right>_{2}^{CB}, \hspace{1cm}				
	\left|GHZ\right>_{23} =\left|{\cal PA}\right>_{3}^{CB}.
\end{align}
The probabilities associated with these measures are given by
\begin{align}
	\mathcal{P}_{20} =\mathcal{P}_{23} =& \dfrac{(NN_m)^2}{8p_2}\left[
	\left(A_{j/2} B_{j/2}^2((a+b)\beta_+ -(a-b)\beta_-)\right)^2+
	\left(A_{j/2}^2B_{j/2}((a+b)\beta_- -(a-b)\beta_+)\right)^2\right],\nonumber
\end{align}
and	
{ \textit{ \begin{align}
			\mathcal{P}_{21} = \mathcal{P}_{22}= &  \dfrac{(NN_m)^2}{8p_2}\left[
			\left(B_{j/2} A_{j/2}^2((a+b)\beta_- +(a-b)\beta_+) \right)^2+
			\left(B_{j/2}^2A_{j/2}((a+b)\beta_+ +(a-b)\beta_-)\right)^2\right].
\end{align}}}
Easily, one can work out the corresponding teleportation fidelity such as
\begin{align}
\mathcal{F}_{20} = \dfrac{(N^2N_m)^2}{2p_2p_{20}}|A_{j/2}^2B_{j/2}^2(a^2\beta_- + 2ab\beta_+ - b^2\beta_+)|^2,\hspace{1cm}\mathcal{F}_{23} =  \dfrac{2(N^2N_m)^2}{p_2p_{23}}|A_{j/2}^2B_{j/2}^2ab(1+e^{im\pi})|^2,\nonumber
\end{align}
\begin{align}
	\mathcal{F}_{21} =  \dfrac{(N^2N_m)^2}{2p_2p_{21}}			|B_{j/2}A_{j/2}^3(a+b)(b-ae^{im\pi})
	-A_{j/2}B_{j/2}^3(a-b)(b+ae^{im\pi})|^2,\nonumber
\end{align}							
\begin{align}
	\mathcal{F}_{22} =  \dfrac{(N^2N_m)^2}{2p_2p_{22}}
	|B_{j/2}A_{j/2}^3(a+b)(b-ae^{im\pi})
	+A_{j/2}B_{j/2}^3(a-b)(b+ae^{im\pi})|^2.\nonumber
\end{align}
Therefore, the average fidelity (\ref{faveragF}) for the first repeated attempt of GHZ states measurements becomes	
\begin{align}	
	\mathcal{F}^{(1)}_{av.2} = \mathcal{F}_0 \mathcal{P}_0 + \mathcal{F}_3 \mathcal{P}_3 + \mathcal{P}_1\left( \mathcal{P}_{10}\mathcal{F}_{10}+\mathcal{P}_{13}\mathcal{F}_{13}\right) + \mathcal{P}_2\left( \mathcal{P}_{20}\mathcal{F}_{20}+\mathcal{P}_{23}\mathcal{F}_{23}\right).
\end{align}
Just as primary attempt, we set $NA_{j/2}(a+b)= \cos(\omega /2)$ and $NB_{j/2}(a-b)= \sin(\omega /2)$, where $\omega \in [0, 2 \pi]$. Then the average fidelity at this stage proved to be
\begin{align}
	F^{(1)}_{av.2} = \dfrac{1}{8}\left(3\left( 1 +|\cos \omega|^2\right) +2
	\sqrt{1-{\cal C}^2} \cos\omega\right).
\end{align}
In the situation where the teleported state is the superposed target state (for $\omega=\pi/2$), the average fidelity is $F^{(1)}_{av.2} = 3/4$. For the separable target state, we have
\begin{equation}
	F^{(1)}_{av.2}=\left\lbrace
	\begin{aligned}
		\dfrac{3+\sqrt{1-{\cal C}^2}}{4},\hspace{1cm}{\rm if}\hspace{1cm}\omega=0,\\
		\dfrac{3-\sqrt{1-{\cal C}^2}}{4},\hspace{1cm}{\rm if}\hspace{1cm}\omega=\pi.
	\end{aligned}
	\right.
\end{equation}
For this attempt, it is easy to show that the average fidelity is minimal when $\omega=\pi/2$ and its minimum value is $\mathcal{F}^{(1)}_{av.min.2}=3/8$. If one studies average and maximizes the angle $\omega = (0;\pi)$, one has
\begin{align}
	\mathcal{F}^{(1)}_{av.max.2} = 2 \mathcal{F}^{(1)}_{av.min.2}+ \dfrac{p^j}{2(1+p^{2j})}.
\end{align}	
It must be noticed that out of the eight cases of the first repeated attempt GHZSMs, Alice can achieve success in four cases. Thus success probability is increased by the quantity
\begin{equation}
	P^{(1)}_{success.2} = P_{success.2}+\mathcal{P}_1\left( \mathcal{P}_{10}+\mathcal{P}_{13}\right) + \mathcal{P}_2\left( \mathcal{P}_{20}+\mathcal{P}_{23}\right),
\end{equation}
due to the product of probability for failure in primary GHZSMs and the sum of probabilities for the results $\left|GHZ\right>_{10}$, $\left|GHZ\right>_{13}$,
$\left|GHZ\right>_{20}$ and $\left|GHZ\right>_{23}$. This remarkable improvement of the success probability due to increasing success cases. At this stage, the success probability takes the form
\begin{equation}
	P^{(1)}_{success.2}  =\dfrac{1}{4}\left( 3 + \dfrac{2p^j cos\omega}{1 + p^{2j}}\right).\label{PS21}
\end{equation}
For $m=0$, the corresponding average fidelity and probabilities after the measurements are reported in Table.(\ref{tab2}). Similarly, for $m=1$, two failure cases of GHZ states measurements occur and are associated with the measurements $\left|GHZ\right>_{0}$ and $\left|GHZ\right>_{3}$ (see table.\ref{tab1}). We consider these cases one by one:

\begin{table}[h!]
	\centering
	\renewcommand{\arraystretch}{1.3}
	\begin{tabular}{|c|c|c|c|}
		\hline \rowcolor{lightgray} GHZSM & State of particle A & $\mathcal{F}{(1)}_{max.2}$ & $\mathcal{P}$ \\
		\hline  $\left|GHZ\right>_{10}$   &  $\begin{array} {lcl}\left|T_{10}\right>=\dfrac{NN_m}{2\sqrt{2p_1p_{10}}}&(&B_{j/2}^2A_{j/2}[(a+b)\beta_+ +(a-b)\beta_-]\left|0\right>^{\otimes j}\\&+&B_{j/2}A_{j/2}^2[(a+b)\beta_{-} + (a-b)\beta_+]\left|1\right>^{\otimes j})_{A}\end{array}$ &  $1$& $\mathcal{P}_1\mathcal{P}_{10}$       \\
		\hline  $\left|GHZ\right>_{11}$&  $\begin{array} {lcl}|T_{11}>=\dfrac{NN_m}{2\sqrt{2p_1p_{11}}}&(&B_{j/2}A_{j/2}^2[(a+b)\beta_{-} +(a-b)\beta_{+}]\left|0\right>^{\otimes j} \\&+&B_{j/2}^2A_{j/2}[(a+b)\beta_{+} + (a-b)\beta_{-}]\left|1\right>^{\otimes j})_{A}\end{array}$&  $\neq 1 $ & $\mathcal{P}_1\mathcal{P}_{11}$  \\
		\hline  $\left|GHZ\right>_{12}$ &$\begin{array} {lcl}\left|T_{12}\right>=\dfrac{NN_m}{2\sqrt{2p_1p_{12}}}&(&B_{j/2}A_{j/2}^2[(a+b)\beta_{-} + (a-b)\beta_{+}]\left|0\right>^{\otimes j} \\&-& B_{j/2}^2A_{j/2}[(a+b)\beta_{+} +(a-b)_{-}]\left|1\right>^{\otimes j})_{A}\end{array}$& $\neq 1$ & $p_1p_{12}$ \\
		\hline    $\left|GHZ\right >_{13}$ &$\begin{array} {lcl}\left|T_{13}\right>=\dfrac{NN_m}{2\sqrt{2p_1p_{13}}}&(&-B_{j/2}^2A_{j}[(a+b)\beta_{+} + (a-b)\beta_{-}]\left|0\right>^{\otimes j}\\&+&B_{j/2}A_{j/2}^2[(a+b)\beta_{-} + (a-b)\beta_{+}]\left|1\right>^{\otimes j})_{A}\end{array}$&  $1$  & $\mathcal{P}_1\mathcal{P}_{13}$ \\					
		\hline $\left|GHZ\right >_{20}$ & $\begin{array} {lcl}\left|T_{20}\right>=\dfrac{NN_m}{2\sqrt{2p_2p_{20}}}&(&A_{j/2}(B_{j/2})^2[-(a+b)\beta_{+} + (a-b)\beta_{-}]\left|0\right>^{\otimes j}\\&+&A_{j/2}^2B_{j/2}[(a+b)\beta_{-} - (a-b)\beta_{+}]\left|1\right>^{\otimes j})_{A}\end{array}$ & $1$& $\mathcal{P}_2\mathcal{P}_{20}$ \\
		\hline $\left|GHZ\right>_{21}$ &$\begin{array} {lcl}\left|T_{21}\right>=\dfrac{NN_m}{2\sqrt{2p_2p_{21}}}&(&A_{j/2}^2B^{j/2}[(a+b)\beta_{-} -(a-b)\beta_{+}]\left|0\right>^{\otimes j}\\&-&A_{j/2}B_{j/2}^2[(a+b)\beta_{+} -  (a-b)\beta_{-}]\left|1\right>^{\otimes j})_{A}\end{array}$ & $\neq 1$  & $\mathcal{P}_2\mathcal{P}_{21}$\\
		\hline $\left|GHZ\right>_{22}$ & $\begin{array} {lcl}\left|T_{22}\right>=\dfrac{NN_m}{2\sqrt{2p_2p_{22}}}&(&B^{j/2}(A^{j/2})^2[(a+b)\beta_{-} - (a-b)\beta_{+}]\left|0\right>^{\otimes j}\\&+&A^{j/2}(B^{j/2})^2[(a+b)\beta_{+} -  (a-b)\beta_{-}]\left|1\right>^{\otimes j})_{A}\end{array}$ & $\neq 1$ & $\mathcal{P}_2\mathcal{P}_{22}$\\
		\hline $\left|GHZ\right>_{23}$ & $\begin{array} {lcl}\left|T_{23}\right>=\dfrac{NN_m}{2\sqrt{2p_2p_{23}}}&(&(A^{j/2})^2B^{j/2}[(a+b)\beta_{-} - (a-b)\beta_{+}]\left|0\right>^{\otimes j}\\& +& A^{j/2}(B^{j/2})^2[(a+b)\beta_{+} -  (a-b)\beta_{-}]\left|1\right>^{\otimes j})_{A}\end{array}$ & $1$  & $\mathcal{P}_2\mathcal{P}_{23}$  \\										
		\hline
	\end{tabular}
	\caption{Various possible GHZ states measurements results in first repeated attempt on particles $B$ and $C$ for $m=0$.}\label{tab2}
\end{table}						
{\bf Case -3:} The input state after primary GHZ states measurements result $\left|GHZ\right>_{0}$ is
\begin{align}
	\left	|\psi{(0)} \right>_{ABC}^{m=1}=\left|GHZ \right>_{0}\otimes \left|T_{0} \right\rangle=	\dfrac{NN_m}{2\sqrt{p_0}}(&\left|0\right>^{\otimes j}_{C}\otimes\left|0\right>^{\otimes j}_{A} + \left|1\right>^{\otimes j}_{C}\otimes \left|1\right>^{\otimes j}_{A})\otimes(	A_{j/2}[A_{j/2}^2(a+b)\beta_+
	\nonumber\\
	&	+ B_{j/2}^2(a-b)\beta_-]\left|0\right>^{\otimes j}_{B}+B_{j/2} [A_{j/2}^2(a+b)\beta_- + B_{j/2}^2 (a-b)\beta_+]\left|1\right>^{\otimes j}_{B}).
\end{align}
Alice may use orthogonal GHZ basis of subsystems $C$ and $B$ as
\begin{align}
	\left|GHZ\right>_{00}=\left|{\cal PA}\right>_{0}^{CB}, \hspace{1cm}
	\left|GHZ\right>_{01}=\left|{\cal PA}\right>_{1}^{CB}, \hspace{1cm}
	\left|GHZ\right>_{02}=\left|{\cal PA}\right>_{2}^{CB} \hspace{1cm}			
	\left|GHZ\right>_{03} =\left|{\cal PA}\right>_{3}^{CB}.
\end{align}
Incidentally, we obtain
\begin{align}
	\mathcal{P}_{00}=\mathcal{P}_{03}= & \dfrac{(NN_m)^2}{8p_0}\left[
	\left(A_{j/2}(A_{j/2}^2 (a+b)\beta_{+} + B_{j/2}^2  (a-b)\beta_{-}) \right)^2+ \left(B_{j/2}(A_{j/2}^2 (a+b)\beta_{-} + B_{j/2}^2 (a-b)\beta_{+})\right)^2\right],\nonumber
\end{align}
and
\begin{align}
	\mathcal{P}_{01}=\mathcal{P}_{02} = &\dfrac{(NN_m)^2}{8p_0}\left[
	\left(B_{j/2} \left(A_{j/2}^2 (a+b)\beta_- + B_{j/2}^2(a-b)\beta_+\right) \right)^2+ \left(A_{j/2}\left(A_{j/2}^2  (a+b)\beta_+ + B_{j/2}^2  (a-b)\beta_-\right)\right) ^2\right].\nonumber
\end{align}
Thus, one can easily work out the following four fidelity
\begin{align}
	\mathcal{F}_{00} & = \dfrac{(N^2N_m)^2}{8p_0p_{00}}|(A_{j/2}^2(a+b) +B_{j/2}^2(a-b))^2 + (A_{j/2}^2(a+b) -B_{j/2}^2(a-b))^2e^{im\pi}|^2,\nonumber
\end{align}
\begin{align}
	\mathcal{F}_{01} & =  \dfrac{(N^2N_m)^2}{2p_0p_{01}}|B_{j/2}A_{j/2}^3(a+b)(a-be^{im\pi})+A_{j/2}B_{j/2}^3(a-b)(a+be^{im\pi})|^2,\nonumber
\end{align}					
\begin{align}
	\mathcal{F}_{02} & =  \dfrac{N^4N_m^2}{2p_0p_{02}}	|B_{j/2}A_{j/2}^3(a+b)(b-ae^{im\pi})	+A_{j/2}B_{j/2}^3(a-b)(b+ae^{im\pi})|^2,\nonumber
\end{align}	
\begin{align}
	\mathcal{F}_{03} & =  \dfrac{N^4N_m^2}{8p_0p_{03}}|(A_{j/2}^4(a+b)^2-B_{j/2}^4(a-b)^2)(1+e^{im\pi}) |^2.\nonumber
\end{align}	

{\bf Case -4:} In this last case, we employ the result of $\left|GHZ\right>_{3}$ where the input state becomes
\begin{align}
	\left|\psi{(3)} \right>_{ABC}^{m=1}=\left|GHZ \right>_{3}\otimes \left|T_{3} \right\rangle=	\dfrac{NN_m}{2\sqrt{p_3}}(&\left|0\right>^{\otimes j}_{C}\otimes\left|0\right>^{\otimes j}_{A} - \left|1\right>^{\otimes j}_{C}\otimes \left|1\right>^{\otimes j}_{A})\otimes(	A_{j/2}[A_{j/2}^2(a+b)\beta_+
	\nonumber\\
	&	- B_{j/2}^2(a-b)\beta_-]\left|0\right>^{\otimes j}_{B}+B_{j/2} [A_{j/2}^2(a+b)\beta_- - B_{j/2}^2 (a-b)\beta_+]\left|1\right>^{\otimes j}_{B}).
\end{align}
Here, Alice may choose orthogonal GHZ basis of subsystems $C$ and $B$ as
\begin{align}
	\left|GHZ\right>_{30} =\left|{\cal PA}\right>_{0}^{CB}, \hspace{1cm}
	\left|GHZ\right>_{31} =\left|{\cal PA}\right>_{1}^{CB},\hspace{1cm}
	\left|GHZ\right>_{32} =\left|{\cal PA}\right>_{2}^{CB}, \hspace{1cm}				
	\left	|GHZ\right> _{33} =\left|{\cal PA}\right>_{3}^{CB}.
\end{align}		
The probabilities of the outcomes are given by
\begin{align}
	\mathcal{P}_{30}= \mathcal{P}_{33} = &\dfrac{(NN_m)^2}{8p_3}\left[
	\left( A_{j/2}(A_{j/2}^2 (a+b)\beta_+ - B_{j/2}^2  (a-b)\beta_-) \right)^2+\left(B_{j/2}(A_{j/2}^2 (a+b)\beta_- - B_{j/2}^2 (a-b)\beta_+)\right)^2\right],\nonumber
\end{align}
and
\begin{align}
	\mathcal{P}_{31}=\mathcal{P}_{32} = &\dfrac{(NN_m)^2}{8p_3}\left[
	\left(B_{j/2} \left(A_{j/2}^2 (a+b)\beta_{-}- B_{j/2}^2(a-b)\beta_{+}\right) \right)^2+ \left(A_{j/2}\left(A_{j/2}^2  (a+b)\beta_+ - B_{j/2}^2  (a-b)\beta_{-}\right)\right)^2\right].
\end{align}
We therefore explicitly derive the teleportation fidelities as follows
\begin{align}
	\mathcal{F}_{30} & = \dfrac{N^4N_m^2}{8p_3p_{30}}	|(A_{j/2}^2(a+b) -B_{j/2}^2(a-b))^2 + (A_{j/2}^2(a+b) +B_{j/2}^2(a-b))^2e^{im\pi}|^2,\nonumber
\end{align}
\begin{align}
	\mathcal{F}_{31} & =  \dfrac{N^4N_m^2}{2p_3p_{31}}	|B_{j/2}A_{j/2}^3(a+b)(b-ae^{im\pi})
	-A_{j/2}B_{j/2}^3(a-b)(b+ae^{im\pi})|^2,\nonumber
\end{align}
\begin{align}
	\mathcal{F}_{32} & =  \dfrac{N^4N_m^2}{2p_3p_{32}}
	|B_{j/2}A_{j/2}^3(a+b)(a-be^{im\pi})
	-A_{j/2}B_{j/2}^3(a-b)(a+be^{im\pi})|^2,\nonumber
\end{align}													\begin{align}
	\mathcal{F}_{33} & =  \dfrac{N^4N_m^2}{8p_3p_{33}}|(A_{j/2}^4(a+b)^2-
	B_{j/2}^4(a-b)^2)(1+e^{im\pi}) |^2.\nonumber
\end{align}
Consequently, for this first repeated attempt of GHZ states measurements based on failure cases when $m=1$, the average fidelity takes the form		
\begin{align}	
	\mathcal{F}^{(1)}_{av.1} & = \mathcal{F}_1 \mathcal{P}_1 + \mathcal{F}_2 \mathcal{P}_2 + \mathcal{P}_0\left( \mathcal{P}_{01}\mathcal{F}_{01}+\mathcal{P}_{02}\mathcal{F}_{02}\right)+\mathcal{P}_3\left(\mathcal{P}_{31}\mathcal{F}_{31}+\mathcal{P}_{32}\mathcal{F}_{32}\right)= \dfrac{3}{8}\left( 1 +|cos \omega|^2 \right).		
\end{align}
From this result, the minimum average fidelity of the teleported state with $m=1$ occurs at the value $\omega=\pi/2$ and it is $\mathcal{F}^{(1)}_{av.min.1}=3/8$. The maximal value of this quantity is given by
\begin{align}
	\mathcal{F}^{(1)}_{av.max.1} = 2\mathcal{F}^{(1)}_{av.min.1},																						\end{align}													
for $\omega = (0,\pi)$. Indeed, the expression of  probability success is
\begin{align}
	P^{(1)}_{success.1} = P_{success.1} + \mathcal{P}_0\left( \mathcal{P}_{01}+\mathcal{P}_{02}\right) + \mathcal{P}_3\left(\mathcal{P}_{31}+\mathcal{P}_{32}\right) =\dfrac{3}{4}.											
\end{align}
The possible results in all the above cases are summarized in Table.(\ref{tab3}). This teleportation scheme is not optimal and we need to make a second repeated attempt of GHZ states measurements.
	\begin{table}[h!]
	\centering
	\renewcommand{\arraystretch}{1.3}
	\begin{tabular}{|c|c|c|c|}
		\hline \rowcolor{lightgray}GHZSM & State of particle A & $\mathcal{F}{(1)}_{max.1}$ & $\mathcal{P}$ \\
		\hline  $\left|GHZ\right>_{00}$   &
		
		$\begin{array} {lcl}\left|T_{00}\right>=\dfrac{NN_m}{2\sqrt{2p_0p_{00}}}&(&A_{j/2}[A_{j/2}^2(a+b)\beta_+ + B_{j/2}^2(a-b)\beta_-]\left|0\right> ^{\otimes j}\\&+& B_{j/2}[A_{j/2}^2(a+b)\beta_- +B_{j/2}^2 (a-b)\beta_+]\left|1\right> ^{\otimes j})_{A}\end{array}$ &  $\neq 1$& $\mathcal{P}_0\mathcal{P}_{00}$\\
		\hline  $\left|GHZ\right>_{01}$&  $\begin{array} {lcl}\left|T_{01}\right>=	\dfrac{NN_m}{2\sqrt{2p_0p_{01}}}&(&B_{j/2}[A_{j/2}^2  (a+b)\beta_- +B_{j/2}^2 (a-b)\beta_+]\left|0\right> ^{\otimes j}\\&+&A_{j/2} [A_{j/2}^2(a+b)\beta_{+} +  B_{j/2}^2(a-b)\beta_-]\left|1\right> ^{\otimes j})_{A}\end{array}$&  $1 $ & $\mathcal{P}_0\mathcal{P}_{01}$ \\
		
		\hline  $\left|GHZ\right>_{02}$ &$\begin{array} {lcl}\left|T_{02
			}\right>=\dfrac{NN_m}{2\sqrt{2p_0p_{02}}}&(&B_{j/2}[ A_{j/2}^2(a+b)\beta_- +B_{j/2}^2 (a-b)\beta_+]\left|0\right>^{\otimes j}\\&-&[A_{j/2}(A_{j/2}^2(a+b)\beta_{+} +B_{j/2}^2  (a-b)\beta_-)]\left|1\right> ^{\otimes j})_{A}\end{array}	$& $1$  & $\mathcal{P}_0\mathcal{P}_{02}$ \\																						\hline	
		$\left|GHZ\right>_{03}$ &$\begin{array} {lcl}\left|T_{03}\right>=\dfrac{NN_m}{2\sqrt{2p_0p_{03}}}&(&
			A_{j/2} [A_{j/2}^2(a+b)\beta_+ + B_{j/2}^2 (a-b)\beta_-]\left|0\right> ^{\otimes j}\\&-&
			B^{j/2} [A_{j/2}^2(a+b)\beta_- +B_{j/2}^2 (a-b)\beta_+]\left|1\right> ^{\otimes j})_{A}\end{array}$&  $\neq 1$ & $\mathcal{P}_0\mathcal{P}_{03}$  \\	
		\hline $\left|GHZ\right>_{30}$ &$\begin{array} {lcl}\left|T_{30}\right>=	\dfrac{NN_m}{2\sqrt{2p_3p_{30}}}&(&
			A_{j/2}	[A_{j/2}^2(a+b)\beta_+ - B_{j/2}^2(a-b)\beta_-]\left|0\right> ^{\otimes j}\\&-&
			B_{j/2}[A_{j/2}^2(a+b)\beta_- -B_{j/2}^2 (a-b)\beta_+]\left|1\right> ^{\otimes j})_{A}\end{array}$ & $\neq 1$& $\mathcal{P}_3\mathcal{P}_{30}$ \\
		\hline $\left|GHZ\right>_{31}$ &$\begin{array} {lcl}\left|T_{31}\right>=\dfrac{NN_m}{2\sqrt{2p_3p_{31}}}&(&B_{j/2}	[A_{j/2}^2  (a+b)\beta_- -B_{j/2}^2 (a-b)\beta_+]\left|0\right> ^{\otimes j}\\&-&	A_{j/2} [A_{j/2}^2(a+b)\beta_+ -  B_{j/2}^2(a-b)\beta_-]\left|1\right> ^{\otimes j})_{A}\end{array}$ & $1$  & $\mathcal{P}_3\mathcal{P}_{31}$\\
		
		\hline $\left|GHZ\right>_{32}$ &$\begin{array} {lcl}\left|T_{32}\right>=\dfrac{NN_m}{2\sqrt{2p_3p_{32}}}&(&	B_{j/2}[ A_{j/2}^2(a+b)\beta_{-} -B_{j/2}^2 (a-b)\beta_+
			]\left|0\right> ^{\otimes j}\\&+& [A_{j/2}(A_{j/2}^2(a+b)\beta_+ -B_{j/2}^2  (a-b)\beta_-)]\left|1\right> ^{\otimes j})_{A}\end{array}$ & $1$ & $\mathcal{P}_3\mathcal{P}_{32}$  \\
		\hline $\left|GHZ\right>_{33}$ &$\begin{array} {lcl}\left|T_{33}\right>=\dfrac{NN_m}{2\sqrt{2p_3p_{33}}}&(&A_{j/2} [A_{j/2}^2(a+b)\beta_{+} - B_{j/2}^2 (a-b)\beta_-]\left|0\right> ^{\otimes j}\\&+&B^{j/2} [A_{j/2}^2(a+b)\beta_{-} -B_{j/2}^2 (a-b)\beta_+]\left|1\right> ^{\otimes j})_{A}\end{array}$ & $\neq 1$  & $\mathcal{P}_3\mathcal{P}_{33}$ \\										
		\hline
	\end{tabular}
	\caption{GHZ states measurements results in first  repeated attempt on particles $B$ and $C$ for $m=1$.}\label{tab3}
\end{table}
\subsection{Second repeated attempt of GHZ states measurements}
Clearly, the same procedure may be applied when we achieve failure in the first attempt of repeated GHZSMs. This time $4$ cases of failure in first attempt of repeated GHZSMs are considered case by case for $m=0$ and $m=1$, respectively.\par
For $m=0$, four failure cases were GHZ states measurements results  $\left|GHZ\right>_{11}$, $\left|GHZ\right>_{12}$, $\left|GHZ\right>_{21}$  and $\left|GHZ\right>_{22}$. In particular, out of $16$ cases of second repeated attempt of GHZ states measurements, success is found in $8$ cases. At this stage, the expression of average fidelity thus obtained as			
\begin{align}
	\mathcal{F}^{(2)}_{av.2}=& \mathcal{F}^{(1)}_{av.2}
	+\mathcal{P}_1 \mathcal{P}_{11}\left( \mathcal{P}_{110}\mathcal{F}_{110}+\mathcal{P}_{113}\mathcal{F}_{113}\right)+ \mathcal{P}_1\mathcal{P}_{12}\left( \mathcal{P}_{120}\mathcal{F}_{120}+\mathcal{P}_{123}\mathcal{F}_{123}\right)\nonumber \\
	&+\mathcal{P}_2 \mathcal{P}_{21}\left(\mathcal{P}_{210}\mathcal{F}_{210}
	+\mathcal{P}_{213}\mathcal{F}_{213}\right)+ \mathcal{P}_2 \mathcal{P}_{22}\left(\mathcal{P}_{220}\mathcal{F}_{220}+\mathcal{P}_{223}\mathcal{F}_{223}\right).
\end{align}
The explicit form of this quantity in terms of concurrence can be expressed as
\begin{align}
	F^{(2)}_{av.2} = F^{(1)}_{av.2}+ \dfrac{1}{32}\left(|\sqrt{1+{\cal C}^2}-\sqrt{1-{\cal C}^2}\cos \omega|^2+|\sqrt{1-{\cal C}^2}-\sqrt{1+{\cal C}^2}\cos\omega|^2\right),
\end{align}
where for the superposed target state ($\omega=\pi/2$), we find $F{(2)}_{av.2}=7/16$, and for the separable target state, we get:
\begin{equation}
	F^{(2)}_{av.2}=\left\lbrace
	\begin{aligned}
		\dfrac{7+\sqrt{1-{\cal C}^2}}{8},\hspace{1cm}{\rm if}\hspace{1cm}\omega=0,\\
		\dfrac{7-\sqrt{1-{\cal C}^2}}{8},\hspace{1cm}{\rm if}\hspace{1cm}\omega=\pi.
	\end{aligned}\label{eq54}
	\right.
\end{equation}
For this second repeated attempt of GHZSMs, we note that the average fidelity $\mathcal{F}^{(2)}_{av.2}$ increases monotonically to reach the unity independently of the spin $j$. In addition, the expression of success probability reads
\begin{align}
	P^{(2)}_{success.2} & = P^{(1)}_{success.2} +\mathcal{P}_1 \mathcal{P}_{11}\left(\mathcal{P}_{110}+\mathcal{P}_{113}\right)+ \mathcal{P}_1\mathcal{P}_{12}\left( \mathcal{P}_{120}+\mathcal{P}_{123}\right)+ \nonumber \\
	&   \mathcal{P}_2 \mathcal{P}_{21}\left( \mathcal{P}_{210} +\mathcal{P}_{213}\right)+ \mathcal{P}_2 \mathcal{P}_{22}\left(\mathcal{P}_{220}+\mathcal{P}_{223}\right)=\dfrac{1}{8}\left(7+ \dfrac{2p^jcos\omega}{1 + p^{2j}} \right).\label{PS22}
\end{align}
Thus, we can say that our scheme gives an almost perfect quantum teleportation of GHZSMs and it becomes effective with higher values of the overllap $p$ (i.e., for large coherent amplitudes) and for all values of $\omega$.\par

For $m=1$, the $4$ failure cases were GHZ states measurements results  $\left|GHZ\right>_{00}$, $\left|GHZ\right>_{03}$, $\left|GHZ\right>_{30}$  and $\left|GHZ\right>_{33}$. Similarly, success is achieved in $8$ cases. At this stage, the expression of average fidelity becomes
\begin{align}
	\mathcal{F}^{(2)}_{av.1} & = \mathcal{F}^{(1)}_{av.1}
	+\mathcal{P}_0 \mathcal{P}_{00}\left( \mathcal{P}_{001}\mathcal{F}_{001}+\mathcal{P}_{002}\mathcal{F}_{002}\right)+ \mathcal{P}_0\mathcal{P}_{03}\left( \mathcal{P}_{031}\mathcal{F}_{031}+\mathcal{P}_{032}\mathcal{F}_{032}\right)
	\nonumber \\
	&+\mathcal{P}_3 \mathcal{P}_{30}\left( \mathcal{P}_{301}\mathcal{F}_{301}
	+\mathcal{P}_{302}\mathcal{F}_{302}\right)	
	+ \mathcal{P}_3 \mathcal{P}_{33}\left(\mathcal{P}_{331}\mathcal{F}_{331}+\mathcal{P}_{332}\mathcal{F}_{332}\right) 
	= \dfrac{7}{6} \mathcal{F}^{(1)}_{av.1},
\end{align}
and the success probability is given by
\begin{align}
	P^{(2)}_{success.1} & = P^{(1)}_{success.1}  +\mathcal{P}_0 \mathcal{P}_{00}\left( \mathcal{P}_{001}+\mathcal{P}_{002}\right)+ \mathcal{P}_0\mathcal{P}_{03}\left( \mathcal{P}_{031}+\mathcal{P}_{032}\right)
	\nonumber \\
	&+ \mathcal{P}_3 \mathcal{P}_{30}\left( \mathcal{P}_{301} +\mathcal{P}_{302}\right) + \mathcal{P}_3 \mathcal{P}_{33}\left( \mathcal{P}_{331}+\mathcal{P}_{332}\right)= \dfrac{7}{8}.
\end{align}
It is obvious that we repeats the same procedure in order to achieve a high success as desired. Thus considering third repeated attempt of GHZ states measurements. At this stage, there are $8$ unsuccessful cases of second repeated attempt corresponding to $32$ possibility of third repeated attempt out of which we get $16$ successful cases. Therefore, the final success probability up to third repeated attempt, for $m=0$, is derived as
\begin{align}
	P^{(3)}_{success.2} = P^{(3)}_{success.1}+\dfrac{1}{8} \dfrac{p^jcos\omega}{1 + p^{2j}},\label{PS23}
\end{align}
and $P^{(3)}_{success.1}=15/16$ for $m=1$. From the success probabilities of the first attempt (\ref{PS21}), second attempt (\ref{PS22}) and third attempt (\ref{PS23}), we can determine the success probability of the $n^{th}$ repeated. We easily obtain
\begin{align}
	P^{(n)}_{success.2} = P^{(n)}_{success.1} +\dfrac{1}{2^{n+1}} \sqrt{1-{\cal C}^2}\cos\omega,\label{eq59}
\end{align}
for the symmetric spin coherent state (mod $2$) while the success probability of the $n^{th}$ repeated attempt for the antisymmetric spin coherent state (mod $1$) takes the form
\begin{align}
	P^{(n)}_{success.1} = \dfrac{2^{n+1}-1}{2^{n+1}},\label{eq60}
\end{align} 
which is independent on the parameter $\omega$ of the target state. If the teleported state is the superposed target state with $\omega=\pi/2$, Eq.(\ref{eq59}) reduces to Eq.(\ref{eq60}) and then the resource connecting Alice and Bob in both the symmetric and the antisymmetric state cases have the same success probability.\par
\begin{figure}[h!]
	{{\begin{minipage}[b]{.5\linewidth}
				\centering
				\includegraphics[scale=0.5]{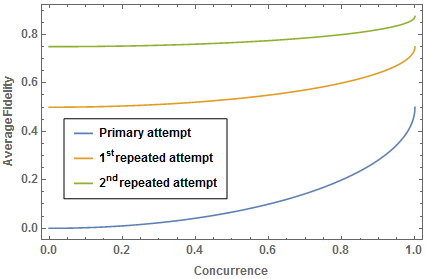}\vfill
				$(a)$
			\end{minipage}\hfill
			\begin{minipage}[b]{.5\linewidth}
				\centering
				\includegraphics[scale=0.5]{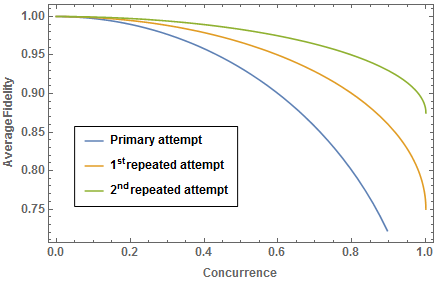}\vfill
				$(b)$
	\end{minipage}}}
	\caption{Variation of the average fidelity with the concurrence of the shared resource state with continuation of GHZSMs when: (a) $m=0$ and $\omega=\pi$, (b) $m=0$ and $\omega=0$.}\label{Fig4}
\end{figure}

The variations of the average fidelity in terms of Wootters concurrence when the shared resource state is a symmetric spin coherent state ($m=0$), for different values taken by the $w$ given in equation (\ref{eq54}), are reported in Figure (\ref{Fig4}). As depicted in Fig.\ref{Fig4}(a), we observe that the average fidelity obtained within the $n^{th}$ repeated attempt of GHZ state measurements (with $n=0,1,2$) displays an exponential behavior between the shared separable resource state (${\cal C}= 0$) and the shared maximally entangled resource state (${\cal C}=1$). Hence, when the target state parameter $\omega=\pi$, the maximum average fidelity is achieved if the resource connecting Alice and Bob is a maximally entangled state. On the contrary, for the target state parameter $w$ is zero as in Fig.\ref{Fig4}(b), we get the opposite behavior of the one reported in Fig.\ref{Fig4}(a), where the maximum average fidelity is attained for the shared separable state and the minimum average fidelity is achieved for the shared maximally entangled state. Based on these results, we can confirm that the maximum average fidelity depends on the parameters of the teleported state which constitutes the information that Alice (sender) will transmit to Bob (receiver). Besides, in both cases and for any teleported state, we find that a higher average fidelity is obtained if GHZ state measurements are performed. This means that by increasing the number of repeated attempts $n$, the value of the average fidelity increases.

\begin{figure}[h!]
	{{\begin{minipage}[b]{.33\linewidth}
				\centering
				\includegraphics[scale=0.4]{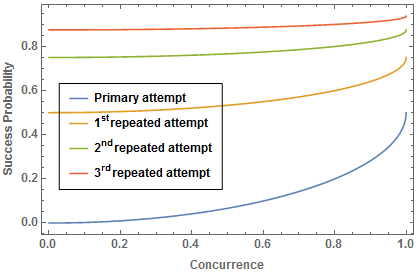}\vfill
				$(a)$
			\end{minipage}\hfill
			\begin{minipage}[b]{.33\linewidth}
				\centering
				\includegraphics[scale=0.4]{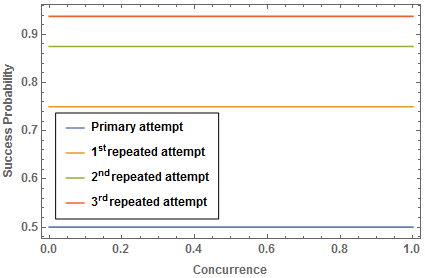}\vfill
				$(b)$
			\end{minipage}\hfill
			\begin{minipage}[b]{.33\linewidth}
				\centering
				\includegraphics[scale=0.4]{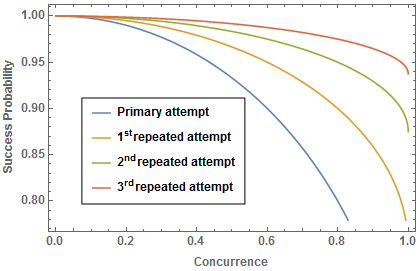}\vfill
				$(c)$
	\end{minipage}}}
	\caption{Improvement of the success probability with the shared resource state entanglement when the teleported state is a separable target state for: (a) the shared resource state is a symmetric spin coherent state with $m=0$ and $\omega=\pi$, (b) the shared resource state is an antisymmetric spin coherent state with $m=1$ and $\omega=0$, (c) the shared resource state is a symmetric spin coherent state with $m=0$ and $\omega=0$.}\label{Fig5}
\end{figure}

In figure (\ref{Fig5}) we have plotted the evolution of the success probability against the shared resource state entanglement when the shared resource state is an symmetric spin coherent state (Fig.\ref{Fig5}(a) and Fig.\ref{Fig5}(c)) and also when it is an antisymmetric spin coherent state (Fig.\ref{Fig5}(b)). We notice that, as expected, the probability success improves by repeating the process of GHZ states measurements. As a matter of fact, as the number of attempts increases, the success probability increases. It is also worthwhile noticing that the behavior of success probability is almost similar to that observed in the average fidelity (see Fig.\ref{Fig4}) and that both are contingent on the target state parameter $w$. Hence we are led to the conclusion that we can achieve perfect teleportation with the highest average fidelity and a high success probability by repeating the GHZ state measurements even if the shared resource state is not a maximally entangled state.

\section{Concluding remarks}			 			
To summarize, we have provided a new scheme involving repeated GHZSMs to improve both the average fidelity and success probability of probabilistic quantum teleportation protocol, using the even and odd spin coherent states as entangled resource. This scheme allows to teleport an unknown state with unit fidelity and high success probability and to transmit quantum information more faithfully using non-maximally entangled states as a preshared quantum resource. Here, Alice initially has all the particles with her and performs repeated GHZ states measurements if she faced with failure, i.e., the maximal fidelity less than $1$. Currently, Alice is looking at more faithful quantum teleportation with probabilistic success, i.e., maximum fidelities in the range of 1, which is significantly higher than the threshold value of 2/3 related to the non-quantum cloning theorem that underlies the security of such quantum communication protocol. In the primary attempt, she faced failure in two cases. That's gives in the first repeated GHZ states measurements four success and four failure cases, these four failure cases gives $8$ success and $8$ failure cases, which gives in the second repeated attempt of GHZ states measurements $16$ success and $16$ failure cases, etc. We show that repeated GHZ states measurements can, in principle, yield a quantum teleportation protocol with maximal fidelity (equal to $1$). This gives more success cases with an improvement of the success probability and teleportation fidelity. In other words, we verify that a higher average fidelity is obtained, then the optimal successful teleportation is determined, if the repeated GHZ state measurements are performed. Therefore, the success probability and average fidelity are accordingly higher.\par 

Based on the variation of the concurrence that gives us for some specific values a maximally entangled shared resource state, a non-maximally entangled shared resource state or a separated shared resource state, we investigated the probabilistic quantum teleportation efficiency of an arbitrary superposed coherent state into each type of these three possible resources. Our study show that even spin coherent state resource gives more interesting average fidelity in comparison with odd spin coherent state. On the other hand, the repetition of GHZ state measurements in this protocol does not affect the security of the quantum teleportation and it remains completely secure. In practice, the number of repetition can be limited due to some reasons. For this case, the choice may be to $(i)$ continue to use the GHZSMs in the $n^{th}$ repetition or $(ii)$ to use the maximal entangled GHZ basis.\par 

With existing technology, this study can stimulate experimental studies to implement the probabilistic quantum teleportation protocol. Our new method can thus be directly applied to other teleported states and for any non-maximally entangled resources. Future work could focus on extending this present probabilistic teleportation scheme to other forms of resources beyond quantum entanglement, such as quantum discord-like \cite{Ollivier2001,Slaoui2018} and quantum coherence \cite{Streltsov2017}. Indeed, these types of quantumness are more robust to decoherence effects for dissipative systems and still go beyond quantum entanglement \cite{Werlang2009,Slaoui2020,Napoli2016}. We hope to report on this subject in a forthcoming work.\\

{\bf Availability of data and material:} Data sharing not applicable to this article as no datasets were generated or analysed during the current study.

\end{document}